# WiMAX Based 60 GHz Millimeter-Wave Communication for Intelligent Transport System Applications

[1]Bera Rabindranath, [2]Sarkar Subir Kumar, [3]Sharma Bikash,   [4]Sur Samarendra Nath, [5]Bhaskar Debasish & [6]Bera Soumyasree

1,3,4,5,6 Sikkim Manipal Institute of Technology, Sikkim Manipal University, Majitar, Rangpo, East Sikkim, 737132:
2. Jadavpur University, Kolkata 700 032.

{rbera50,samar.sur,debasishbhaskar,soumyasree.bera}@gmail.com

*Abstract*

*With the successful worldwide deployment of 3rd generation mobile communication, security aspects are ensured partly. Researchers are now looking for 4G mobile for its deployment with high data rate, enhanced security and reliability so that world should look for CALM, Continuous Air interface for Long and  Medium range communication. This CALM will be a reliable high data rate secured mobile communication  to be deployed for  car to car communication (C2C) for safety application. This paper reviewed the WiMAX ,& 60 GHz RF carrier for C2C. The system is tested at SMIT laboratory with multimedia transmission and reception. With proper deployment of this 60 GHz system on vehicles, the existing commercial products for 802.11P will be required to be replaced or updated soon .*

**Key words:** C2C, CALM , WiMAX, WiFi, VSG, RTSA .

## 1   Introduction

Safety and security are very important in car-to-car communication. It is even more important when wireless systems are used because it is generally perceived that wireless systems are easier to attack than wireline systems. In search of best, secured and reliable communication technology  towards next generation e-car safety application, IEEE 802.16, an emerging wireless technology for deploying broadband wireless metropolitan area network (WMAN), is one the most promising wireless technology for the next-generation ubiquitous network. Though IEEE802.11P WiFi based  products are commercially available for same functionality. But, disadvantages incurred in the Wi-Fi security have been addressed into the IEEE 802.16 standard and also flexibility parameters are also addressed in WiMAX. WiMAX is designed to deliver next-generation, high-speed mobile voice and data services and wireless "last-mile" backhaul connections [1]

The University of Texas at Austin.  IEEE 802.16e (Mobile WiMax) deals with the Data Link Layer security. The Data-Link Layer authentication and authorization makes sure that the network is only accessed by permitted users while the encryption ensures privacy and protects traffic data from hacking or spying by unauthorized users. The WiMAX 802.16e provides number of advanced security protections including: strong traffic encryption, mutual device/user authentication, flexible key management protocol, control/ management message protection, and security protocol optimizations for fast handovers when users switch between different networks. Fig.1 shows a WiMAX architectural components

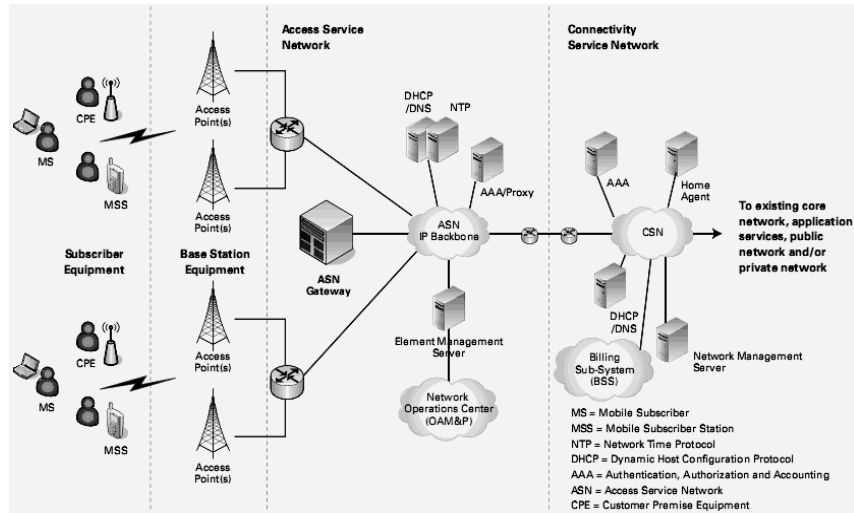

**Fig. 1.** WiMAX architectural components [2].

Commercial products of vehicular networks exists viz. DENSO's Wireless Safety Unit (WSU), Hitachi-Renesas.

DENSO's Wireless Safety Unit (WSU) is the follow up development to DENSO's first generation 802.11p communication module, the Wave Radio Module (WRM). It is specifically designed for automotive environments (temperatures, shock, vibration) and has its primary focus on safety related applications. [4]

During normal driving, the equipped vehicles anonymously share relevant information such as position, speed and heading. In a C2C environment message authorization is vital. The possibility to certify attributes and bind those to certain vehicles is particularly important for public safety. [5]

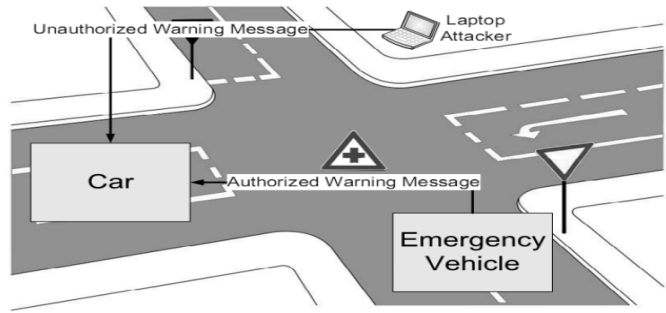

**Fig. 2.** Shows a possible attack in a typical Car2Car environment. Assuming no security, his attacker could generate valid messages for and consequently disturb the whole transportation system.

While unlicensed spectrum around 2.5 GHz and 5 GHz is also available internationally, the amount of available 60G bandwidth is much higher than that around 2.5GHz and 5GHz [3].

Unlicensed spectrum surrounding the 60 GHz carrier frequency has the ability to accommodate high-throughput wireless communications. It is highly directive and can be used for long and directed link. 60GHz system enjoys the size reduction and cost reduction advantages. Additionally, due to availability of 5GHz bandwidth the data-rate for communication is more interesting. Many commercial products have been developed facing these challenges.

Thus ,exploring the WiMAX 802.16e for its security, reliability and high throughput features , exploring 60 GHz millimeter wave as carrier for its size and cost reduction, wide bandwidth and highest throughput , the Car2Car communication system is required to be developed for the next generation Car for safety applications. The development of 60 GHz C2C communication system comprised of  two step procedure discussed below. The MATLAB/SIMULINK is used for the design verification and simulation at the $1^{st}$ stage. The final simulation result in the form of *.mdl file  is ported to the ARB unit of one R&S VSG for the realization of the base band hardware. The transmit IF at 1 GHz and transmit RF at 60 GHz are realized through RF block of R&S VSG and separate 60 GHz transmit module respectively as shown in figure 8 and  9 respectively. The signal reception is developed using 60 GHz RF front end, RF Tuner  and RTSA ( Tektronix real Time spectrum analyzer). in Laptop, Data  is  retrieved at the Laptop  using  data acquisition and digital signal processing.

The above system development efforts are discussed below. Section 2 relates the mathematical modeling of sub-carrier generation. Section 3 will discuss all about WiMAX  simulation at the base band level. The successful development of section 3 will produce one *.mdl file which is ported to the VSG for base band hardware realization.  Section 4 will discuss the efforts pertaining to hardware development.

## 2   Mathematical modeling of Sub-carrier generation

The serial data input is a sequence of samples occurring at interval $T_s$. At the transmitter as shown in Fig 3, the high rate serial input data is converted to column data which is of lower rate. The low-rate column data consists of M low-rate parallel streams in order to increase the symbol duration to $T=MT_s$ . The low-rate streams, represented by the symbols $b_m[k]$, m=0, 1, 2, ........., M-1, and k=1, 2, 3......, where each stream is modulated onto different sub-carriers. The orthogonal relationship between any two of the sub-carriers in a set is maintained to avoid the Inter Channel Interference [8]. Then the parallel streams are multiplexed and a Cyclic Prefix is inserted to eliminate the effect of Inter Symbol Interference. So, we obtain the transmitted $k^{th}$ symbol as below.

$$y(t) = \sum_{m=0}^{M-1} b_m[k]e^{j2\pi mt/T}, \text{ where, } -G + kT \leq t \leq (k+1)T \qquad (1)$$

G is the length of the Guard Interval and at the $m^{th}$ stream of data the kth symbol is coming out as $b_m[k]$.

Now this y(t) will pass through the channel which can be modeled as the frequency selective fading channel and it will also have the multipaths. This type of channel model can be realized if we consider the tapped-delay line with time-varying coefficients with a fixed tap spacing. The following is the mathematical realization of such type of channel [8].

$$h(t,\tau) = \sum_{l=0}^{\chi} h_l(t)\delta(\tau - \tau_l) \qquad (2)$$

$h_l(t)$ and $\tau_l$ are the complex amplitude and delay of the $l^{th}$ path, respectively. $\chi+1$ is the total number of taps. $\tau_\chi$ is the maximum multipath delay spread. For OFDM symbolization, the length of G should be greater than $\tau_\chi$. The expectation value of $h_l(t)$ is the determining factor to get the following correlation function. As $h_l(t)$ is random in nature, it is modeled as the Wide-Sense Stationary Uncorrelated Scattering process.

$$\phi_h(\Delta t) \stackrel{\Delta}{=} E[h_l(t)h_l^*(t - \Delta t)] = \sigma_l^2 \phi_t(\Delta t) \qquad (3)$$

The received signal r(t) in the kth symbol can be represented as

$$r(t) = \int h(t,\tau)y(t-\tau) = \sum_{l=0}^{\chi} h_l(t)y(t-\tau_l) + n(t) \qquad (4)$$

n(t) is the background noise.

For practical implementation, modulation and demodulation can be achieved by Inverse Fast Fourier Transform (IFFT) and Fast Fourier Transform (FFT), respectively. Channel estimation is applied to obtain the estimates of channel fading in each sub-carrier such that coherent detection is achieved. Let q be the sub-carrier index at the output of the OFDM demodulator in a WiMAX system and $s_{k,q}$ is the output for the $q^{th}$ sub-carrier in the $k^{th}$ symbol interval. Now, let us assume that the channel impulse response is quasi-static during the $k^{th}$ symbol interval so that $h(t) \approx h(kT)$ for $kT \leq t < (k+1)T$, the Inter Carrier Interference can be neglected compared to the background noise n(t). Thus the $k^{th}$ sub-carrier output, $s_{k,q}, q \in \{0, 1, 2, ...., M-1\}$, from the demodulator can be expressed as

$$\begin{aligned} s_{k,q} &= \frac{1}{T}\int_{kT}^{(k+1)T}\left[\sum_{l=0}^{\chi} h_l(kT)\times\sum_{m=0}^{M-1} b_m[k]e^{j2\pi m(t-\tau_l)/T} + n(t)\right]e^{-j2\pi qt/T} \\ &= \frac{1}{T}\sum_{l=0}^{\chi} h_l(kT)\sum_{m=0}^{M-1} b_m[k]e^{-j2\pi m\tau_l)/T} \times \int_{kT}^{(k+1)T} e^{j2\pi(m-q)t/T}dt + \frac{1}{T}\int_{kT}^{(k+1)T} n(t)e^{-j2\pi qt/T}dt \\ &= c_{k,q}H_{k,q} + v_{k,q} \end{aligned} \qquad (5)$$

where,

$$c_{k,q} = b_q[k], \; H_{k,q} = \sum_{l=0}^{\chi} h_l(kT)e^{-j2\pi q\tau_l/T} \text{ and } v_{k,q} = \frac{1}{T}\int_{kT}^{(k+1)T} n(t)e^{-j2\pi qt/T}dt \qquad (6)$$

If the channel fading is characterized by $H_{k,q}$ were known, then coherent detection and optimum diversity combining would be achievable at the receiver. $H_{k,q}$ is time varying and usually unknown. Hence, for the accurate estimation of the channel fading parameters, $H_{k,q}$ is to be evaluated for a given $s_{i,q}$, $i \leq k$ and $q = 0, 1, ....., M-1$.

## 3  WiMax Simulation at baseband level

The full WiMAX simulation is shown in the Fig 3.

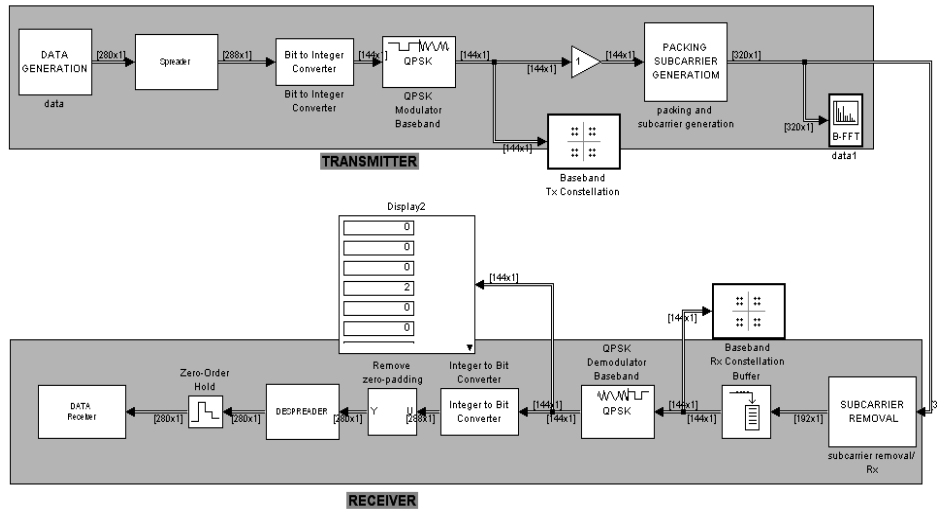

**Fig. 3.** The Simulink model of WiMAX transmitter and receiver.

The descriptions of some important blocks are as follows:

### (1) Data generator [MAC PDU] block under 'data' block:

The 802.16 standard is designed with the network and communication security keeping in considerations. Over the wireless link the robust security aspects are to be considered as the most important to control the confidentiality during data communication. In 802.16 standards, the security keys and encryption techniques are involved as shown in the Fig. 4. It has the similarity in concepts of adopting the security parameters as of IPsec. After the authorization from Security Association, the X.509 certificate, consists of an authorization key (AK), a key encryption key (KEK), and a hash message authentication code (HMAC) key, which are used for authorization, authentication, and key management [6] Here in the following model, we have 10 blocks utilized for message authentication and security management. From the top to bottom, those blocks are: (i) HT: Header Type, (ii) EC: Encryption Control, (iii) Type: Payload Type, (iv) RSV: Reserved, (v) CI: CRC Identifier, (vi) EKS: Encryption Key Sequence, (vii) RSV: Reserved, (viii) LEN: Length of Packet, (ix) CID: Connection Identifier and (x) HCS: Header Check Sequence.

   In terms of message authentication, there are some important shortcomings in IEEE 802.16 standard implemented at the MAC Layer. To avoid the serious threats arise from its authentication schemes, the WiMAX involves a two-way sequential transactions for controlling, authorization and authentication. During the basic and primary connection, MAC management messages are sent in plain text format which is not a robust type of authentication and so can be easily hijacked over the Air and this can be done by the attacker once again. So, as per the X.509 Certificate, the Public Key Infrastructure (PKI) defines a valid connection path to identify a genuine Security Systems. It uses RSA Encryption with SHA1 hashing. The certificate, as pre-configured by the specific manufacturer and embedded within the system must be kept secret so that it can not be stolen by other users/vendors. A Security System that is certificated by a particular manufacturer is implemented in a Base Station (BS) and the particular BS can not know the internal standards priorly.

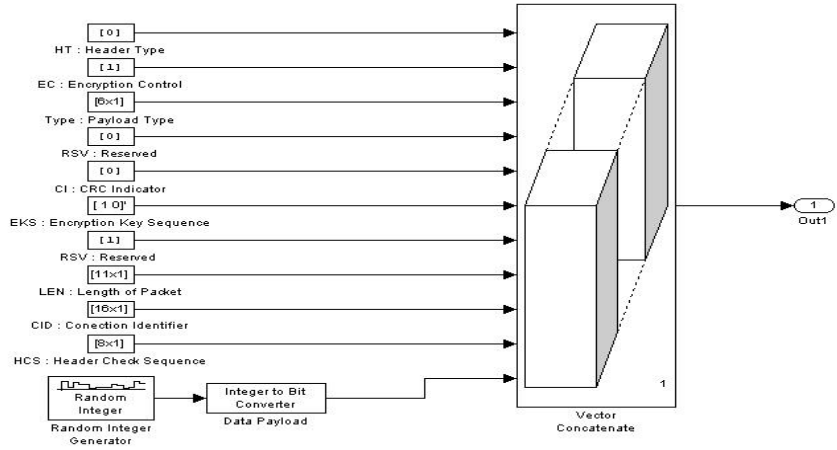

**Fig. 4.** The MAC PDU generator including header scheme and payload

Since, mutual authentication verifies the genuineness of a BS, it should be present in any wireless communication as it is virtually open to all. Extensible Authentication Protocol (EAP) is mostly utilized in any WiMAX Base Stations as to protect IEEE 802.16 / WiMAX against masquerading parties.

The spectrum of the base band just after the Mac PDU packing is shown in Fig.5.

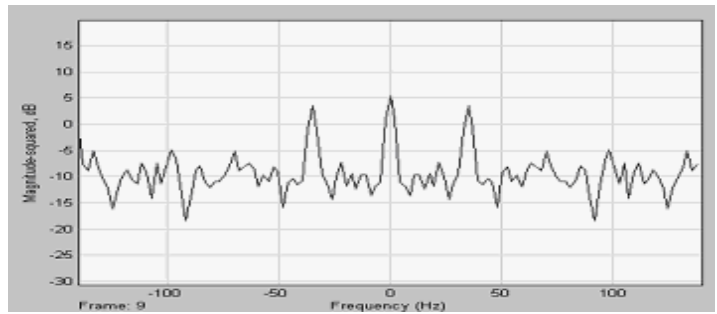

**Fig. 5.** Base band spectrum

### (2) The PN spreading block

The incoming signal is XOR'd with the bit pattern generated by a PN Sequence Generator. This is further zero padded to increase the frame size to 288×1. The Chip sample time is 1/1000 S, so the chip rate is 1 KHz. The spectrum after spreading looks like as shown in the figure 6 below: The bit pattern generated after this is fed into a bit to integer converter and then to a QPSK modulator.

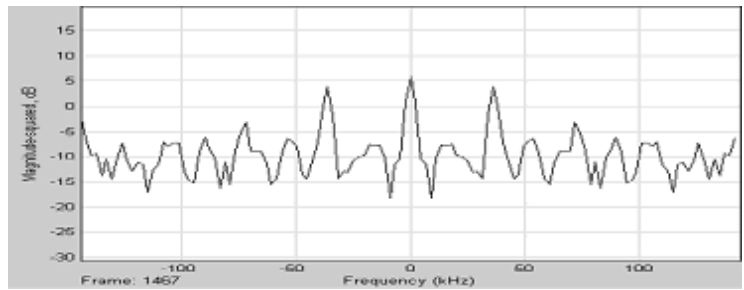

**Fig. 6.** Spectrum after spreading

**(3) The Sub carrier Generation and packing sub system:**

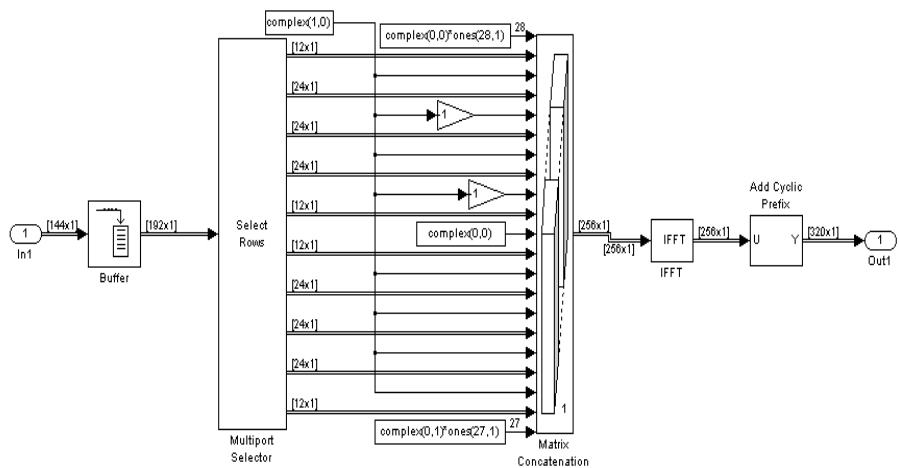

**Fig. 7.** Input packing before IFFT transformation and cyclic prefixing

After the QPSK modulation, pilot is inserted which helps in channel estimation. Here 192×1 input stream is broken down into 10 different data pipes and pilots inserted in between them according to the above figure. All the rows of the resulting data are combined before feeding it to the IFFT block for sub carrier generation in time domain and then cyclic prefixing to add guard time. The final Tx blocks looks like as depicted in Fig.7.

## 4   Hardware Implementation of the 60 GHz System

### 4.1   Description of Transmitter Section

The prototype model of the 60 GHz transmitter is shown in Fig.8 and its block schematic diagram is shown in Fig.9.The Tx section consists of several parts as shown in Fig.9. The PC is used to programme the VSG using Matlab/ Simulink for the generation of two orthogonal basis functions [9]. In the base band section we programmed the ARB section of the VSG to generate the base band WiMAX signal and it is then up converted to IF level of 1 GHz and fed to the 60 GHz varactor tuned Gunn oscillator. The basic block diagram is shown in the Fig.9. The Gunn oscillator is followed by 60 GHz attenuator and frequency meter for the control and frequency measurement of 60 GHz transmitted signal respectively [7]. The 2 feet parabolic disc antenna is connected at the output for radiation of 60 GHz signal

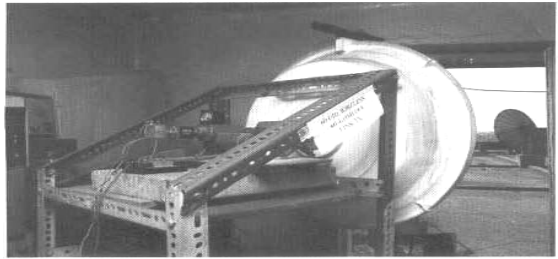

**Fig. 8.** The 60 GHz Transmitter

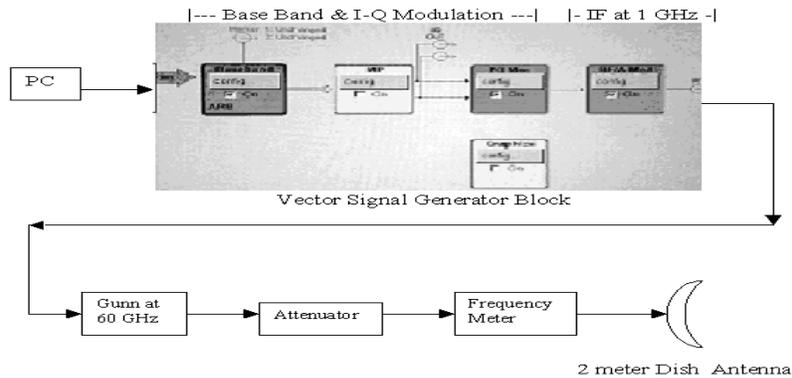

**Fig. 9**. Block diagram of the WiMAX Transmitter

### 4.2 Description of Receiver Section

The prototype model of the 60 GHz receiver is shown in Fig.10. The block schematic diagram is shown in Fig.11 where the receiver consists of a front end, which receives signal through a horn antenna. The received signal then down converted to IF level (1 GHz) using 61GHz Gunn oscillator. This signal is further amplified by two IF amplifiers and is fed to input of the DVB satellite receiver tuner. The I-Q signal from the receiver tuner is connected to the RTSA as shown in Fig.11. We store the received I-Q data in RTSA for further analysis, as shown in Fig.14.

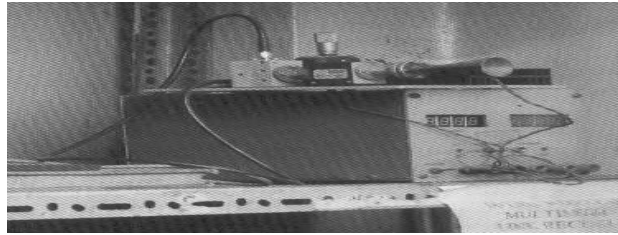

**Fig. 10.** The 60 GHz Received RF Front End

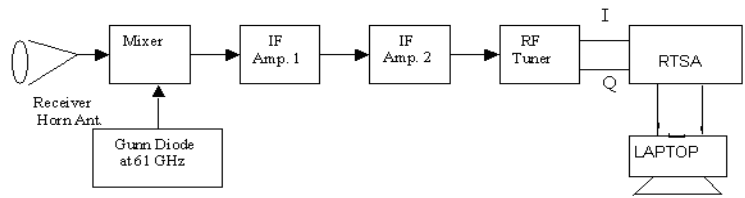

**Fig. 11.** Block diagram of the 60 GHz WiMAX receiver

The Rx spectrum at IF level is shown in Fig.13.

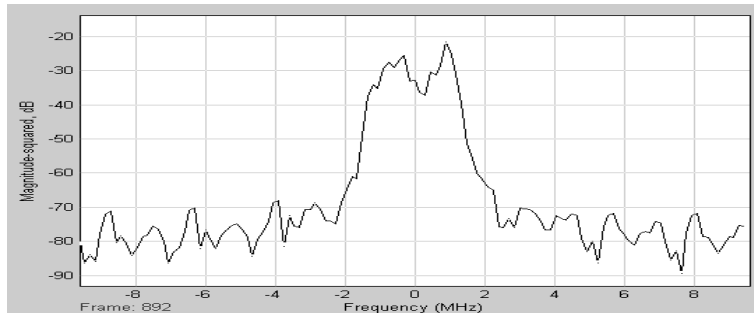

**Fig. 12.** The final transmit spectrum

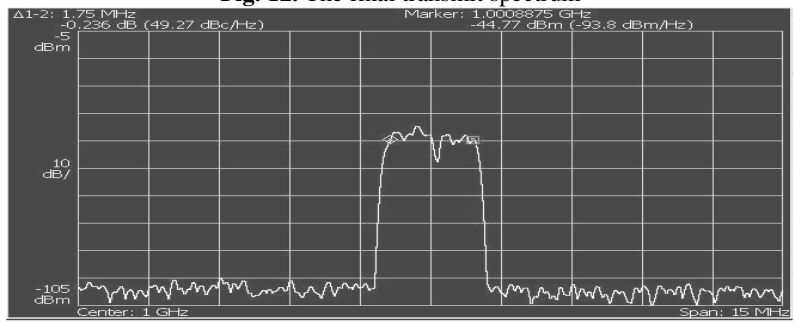

**Fig. 13.** Received spectrum with bandwidth is 1.75 MHz

The Received In-Phase and Quadrature-Phase Signals in Real Time Spectrum Analyzer is shown in figure 14.

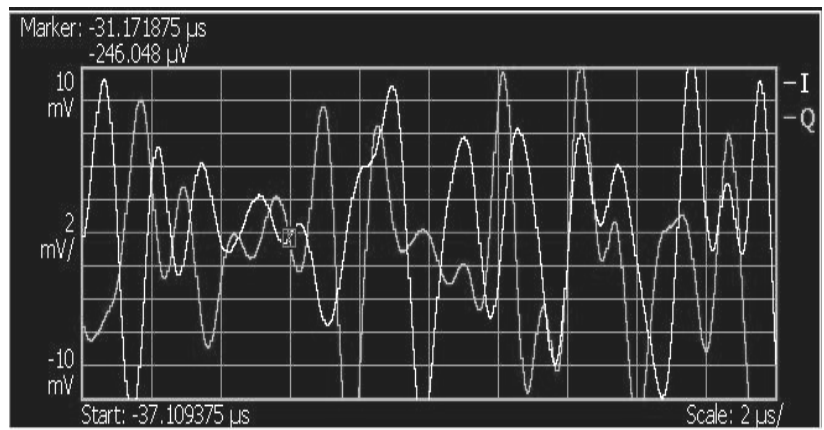

**Fig. 14.** Received I, Q Signals in RTSA

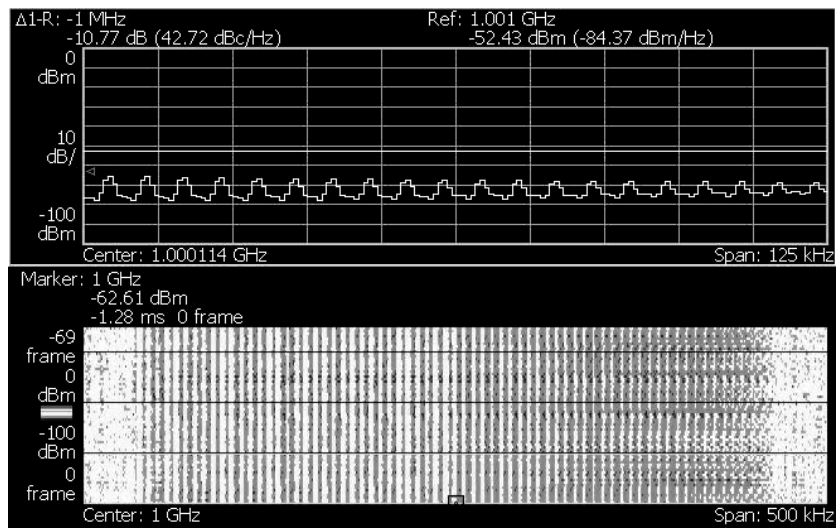

**Fig. 15.** Received WiMAX Sub-Carriers in RTSA

## 5   Conclusion

Lots of efforts are imparted for the development of the 60 GHz C2C link. The system is tested at SMIT laboratory with multimedia transmission and reception. Technical expertise are developed towards Simulink programming, methods of poring to VSG, IF and millimeter wave hardware, RTSA use, Data Acquisition and DSP. The system is operational at SMIT laboratory but yet to be tested after mounting on the vehicles. This successful development encourages the active groups at the laboratory. With proper deployment of this 60 GHz system on vehicles, the existing commercial products for 802.11P will be required to be replaced or updated soon and we look forward for the improved society with intelligent vehicles.